\documentclass%%%%[
  {article}
\usepackage{graphicx}
\newcommand{\lora} {\boldmath$\longrightarrow$}
\newcommand{\vecbm}[1]{\mbox{\boldmath#1}}

\newcommand{\mlora} {\bf\longrightarrow}

\title{ A New Thermodynamics, From Nuclei to Stars}

\author{D.H.E. Gross\\
Hahn-Meitner Institute and Freie Universit{\"a}t Berlin,\\ Fachbereich
Physik.\\ Glienickerstr. 100\\ 14109 Berlin, Germany \footnote{contributed
paper for the international conference "Gobal Phase Diagrams", Odessa
2003}}

\begin{document}

\maketitle

\begin{abstract}
  Equilibrium statistics of Hamiltonian systems is correctly described
  by the microcanonical ensemble. Classically this is the manifold of all
  points in the $N-$body phase space with the given total energy.
  Due to Boltzmann's principle, $e^S=tr(\delta(E-H))$, its
  geometrical size is related to the entropy $S(E,N,\cdots)$. This
  definition does not invoke any information theory, no
  thermodynamic limit, no extensivity, and no homogeneity
  assumption, as are needed in conventional (canonical) thermo-statistics.
  Therefore, it describes the equilibrium statistics of extensive
  as well of non-extensive systems. Due to this fact it is the {\em general
  and
  fundamental} definition of any classical equilibrium statistics. It can
  address nuclei and astrophysical objects as well. As these are not described by
  the conventional extensive Boltzmann-Gibbs thermodynamics, this is a mayor
  achievement of statistical mechanics. Moreover, all kind of phase transitions can be
  distinguished sharply and uniquely for even small systems. In contrast to the
  Yang-Lee singularities in Boltzmann-Gibbs canonical thermodynamics
  phase-separations are well treated.
\end{abstract}

%%%%%%%%%%%%%%%%%%%%%%%%%%%%%%%%%%%%%%%%%%%%
%% MAINMATTER
%%%%%%%%%%%%%%%%%%%%%%%%%%%%%%%%%%%%%%%%%%
\section{Introduction}
Classical Thermodynamics and the theory of phase transitions of
homogeneous and large systems are some of the oldest and best
established theories in physics. It may look strange to add
anything new to this. Let me recapitulate what was told to us
since $>100$ years:
\begin{itemize}
\item Thermodynamics addresses large homogeneous systems at equilibrium
(in the thermodynamic limit
$N\to\infty|_{ N/V=\rho, homogeneous}$).
\item Phase transitions are the positive zeros of the grand-canonical
partition sum $Z(T,\mu,V)$ as function of $e^{\beta\mu}$
(Yang-Lee-singularities). As the partition sum for a finite number
of particles is always positive, zeros can only exist in the
thermodynamic limit $V|_{\beta,\mu}\to\infty$.
\item Micro and canonical ensembles are equivalent.

\item Thermodynamics works with intensive variables $T,P,\mu$.
Fluctuations can mostly be ignored.
\item Unique Legendre mapping $E\Leftrightarrow T$.
\item The heat capacity is always $C>0$.
\item Heat only flows from hot to cold (Clausius)
\item Second Law only in infinite systems when the Poincarr\'{e}
recurrence time becomes infinite (much larger than the age of the
universe (Boltzmann)).
\end{itemize}

 Under these constraint only a tiny part of the real world of
equilibrium systems is addressed. The ubiquitous non-homogeneous
systems: nuclei, clusters, polymer, soft matter (biological)
systems, but even the largest, astrophysical systems are not
covered. Even normal systems with short-range coupling at phase
separations are non-homogeneous and can only be treated within
conventional homogeneous thermodynamics (e.g. van-der-Waals
theory) by bridging the unstable region of negative
compressibility by an artificial Maxwell construction. Thus even
the original goal, for which Thermodynamics was invented some
$150$ years ago, the description of steam engines is only
artificially solved. There is no (grand-) canonical ensemble of
phase separated and, consequently, non-homogeneous,
configurations. This has a deep reason as I will discuss below:
here the systems have a {\em negative} heat capacity (resp.
susceptibility). This, however, is impossible in the
(grand-)canonical theory where $C\propto (\delta E)^2$

\section{Boltzmann's principle}
   The Microcanonical ensemble is the ensemble (manifold)
   of all possible points in the $6N$ dimensional phase space at
   the prescribed sharp energy $E$:
\begin{eqnarray*}
W(E,N,V)&=&\epsilon_0 tr\delta(E-H_N)\\
tr\delta(E-H_N)&=&\int{\frac{d^{3N}p\;d^{3N}q}{N!(2\pi\hbar)^{3N}}
\delta(E-H_N)}.
\end{eqnarray*}
Thermodynamics addresses the whole ensemble. It is ruled by
  the topology of the geometrical size $W(E,N,\cdots)$,
 Boltzmann's principle:
 \begin{equation}
\fbox{\fbox{\vecbm{S=k*lnW}}}
\end{equation}
is the most fundamental definition of the entropy $S$.  Entropy
and with it micro-canonical thermodynamics has therefore a pure
mechanical, {\em geometrical} foundation. No information
theoretical formulation is needed. Moreover, in contrast to the
canonical entropy, $S(E,N,..)$ is everywhere single valued and
multiple differentiable. There is no need for extensivity, no need
of concavity, no need of additivity, and no need of the
thermodynamic limit. This is a great advantage of the geometric
foundation of equilibrium statistics over the conventional
definition by the Boltzmann-Gibbs canonical theory. However,
addressing entropy to finite eventually small systems we will face
a new problem with Zermelo's objection against the monotonic rise
of entropy, the Second Law, when the system approaches its
equilibrium. This is discussed elsewhere \cite{gross183,gross192}.
A second difficulty is: Without the thermodynamic limit surface
effects are not scaled away. However, this is important as the
creation of inhomogeneities and interphase surfaces is on the very
heart of phase transitions of first order.
\section{Topological properties of $S(E,\cdots)$}

The topology of the Hessian of $S(E,\cdots)$, the determinant of curvatures
of $s(e,n)$ determines completely all kinds phase transitions. This is
evidently so, because $e^{S(E)-E/T}$ is the weight of each energy in the
canonical partition sum at given $T$. Consequently, at phase separation
this has at least two maxima, the two phases. And in between two maxima
there must be a minimum where the curvature of $S(E)$ is positive. I.e. the
positive curvature detects phase separation. This is of course also in the
case of several conserved control parameters.
\begin{eqnarray}
d(e,n)&=&\left\|\begin{array}{cc} \frac{\partial^2 s}{\partial
e^2}& \frac{\partial^2 s}{\partial n\partial e}\\ \frac{\partial^2
s}{\partial e\partial n}& \frac{\partial^2 s}{\partial n^2}
\end{array}\right\|=\lambda_1\lambda_2 \label{curvdet}\\
\lambda_1&\ge&\lambda_2\hspace{1cm}\mbox{\lora eigenvectors
:}\hspace{1cm} {\boldmath\vecbm{$v$}_1,\vecbm{$v$}_2}\nonumber
\end{eqnarray}
Of course for a finite system each of these maxima of
$S(E,\cdots)-E/T$ have a non-vanishing width. There are intrinsic
fluctuations in each phase.
\subsection{ Unambiguous signal of phase transitions in a "Small"
system} 
\begin{figure}[h]
\begin{minipage}[h]{8cm}
\includegraphics*[bb =0 0 290 180, angle=0, width=8cm,
clip=true]{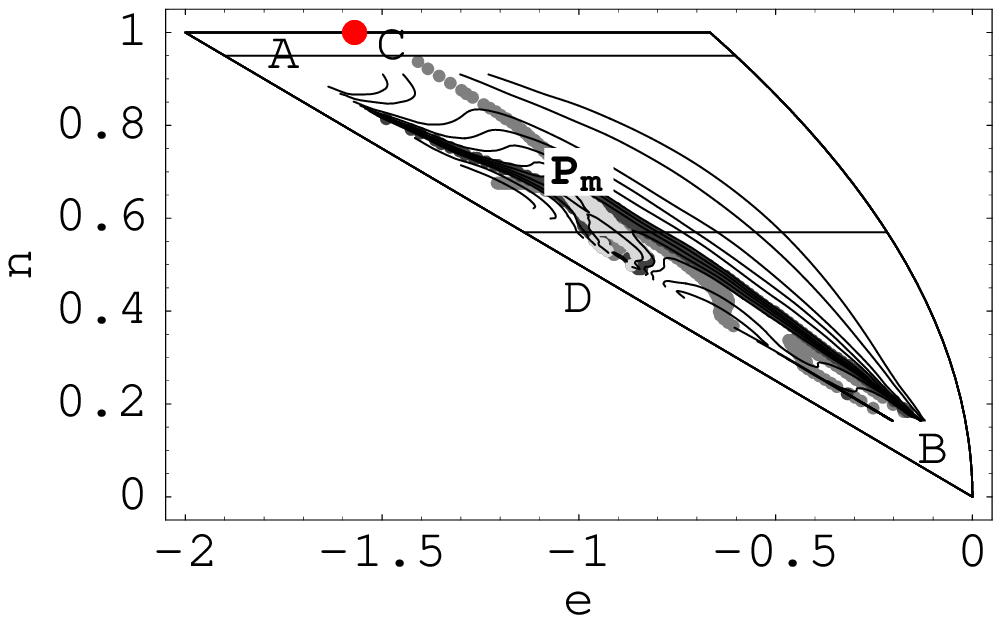}
%%%\includegraphics*[bb =0 0 290 180,
%%%angle=-0, width= 8cm,
%%%clip=true]{p3detg3.eps}%{P3detgrdetcont.eps}~
%%%\end{figure}
\caption{Global phase diagram or contour plot of the curvature
determinant
  (Hessian), eqn.~(\ref{curvdet}), of the 2-dim Potts-3 lattice gas
  with $50*50$ lattice points, $n$ is the number of particles per
  lattice point, $e$ is the total energy per lattice point. The line
   (-2,1) to (0,0) is the ground-state energy of the lattice-gas
   as function of $n$. The most right curve is the locus of configurations
    with completely random spin-orientations (maximum entropy). The whole
    physics}\label{det}
\end{minipage}~~\begin{minipage}[h]{8cm}of the model plays between these
two boundaries.  At the
    dark-gray lines the Hessian is $\det=0$,this is the boundary  of the
    region of phase separation (the triangle $AP_mB$) with a negative
    Hessian ($\lambda_1>0,\lambda_2<0$). Here, we have
    Pseudo-Riemannian geometry. At the light-gray lines
    is a minimum of $\det(e,n)$ in the direction of the largest
    curvature (\vecbm{v}$_{\lambda_{max}}\cdot$\vecbm{$\nabla$}$\det=0$) and
    $\det=0$,these are lines of second order transition. In the triangle
    $AP_mC$ is the pure ordered (solid) phase ($\det>0, \lambda_1<0$).
    Above and right of the line $CP_mB$ is the pure disordered (gas) phase
    ($\det>0, \lambda_1<0$). The crossing $P_m$ of the boundary lines is a
    multi-critical point. It is also the critical end-point of the region
    of phase separation ($\det<0,\lambda_1>0,\lambda_2<0$).  The light-gray
    region around the multi-critical point $P_m$ corresponds to a flat,
    horizontal region of $\det(e,n)\sim 0$ and consequently to a somewhat
    extended cylindrical region of $s(e,n)$ and {\boldmath
    \vecbm{$\nabla$}\mbox{$\lambda_1$}{\boldmath$\sim 0$}},
    details see \protect\cite{gross173,gross174}; $C$ is the analytically
   known position of the critical point which the ordinary $q=3$ Potts
   model (without vacancies){\em would have in the thermodynamic
   limit}
\end{minipage}
\end{figure}

Nevertheless, the whole zoo of phase-transitions can be sharply
seen and distinguished. This is here demonstrated for the Potts-gas model
on a two dimensional lattice of {\em finite} size of $50\times 50$ lattice
points, c.f. fig.(\ref{det}).
%%%\newpage

\subsection{Heat can flow from cold to hot}
Clausius' first version of the second law is violated in regions of
negative heat capacity. Taking energy out of such a system its temperature
$T=(dS/dE)^{-1}$ can rise whereas the temperature of the -- originally--
hotter recipient will drop.

\section{ Negative heat capacity as signal for a phase
transition of first order.}
\subsection{Nuclear Physics}
A very detailed illustration of the appearance of negative heat
capacities in nuclear level densities is given by d.Agostino et
al. \cite{dAgostino00}.
\subsection{Atomic clusters}
\begin{figure}[h]
\begin{minipage}[h]{8cm}
\includegraphics [bb = 69 382 545
766, angle=-0, width=8cm, clip=true]{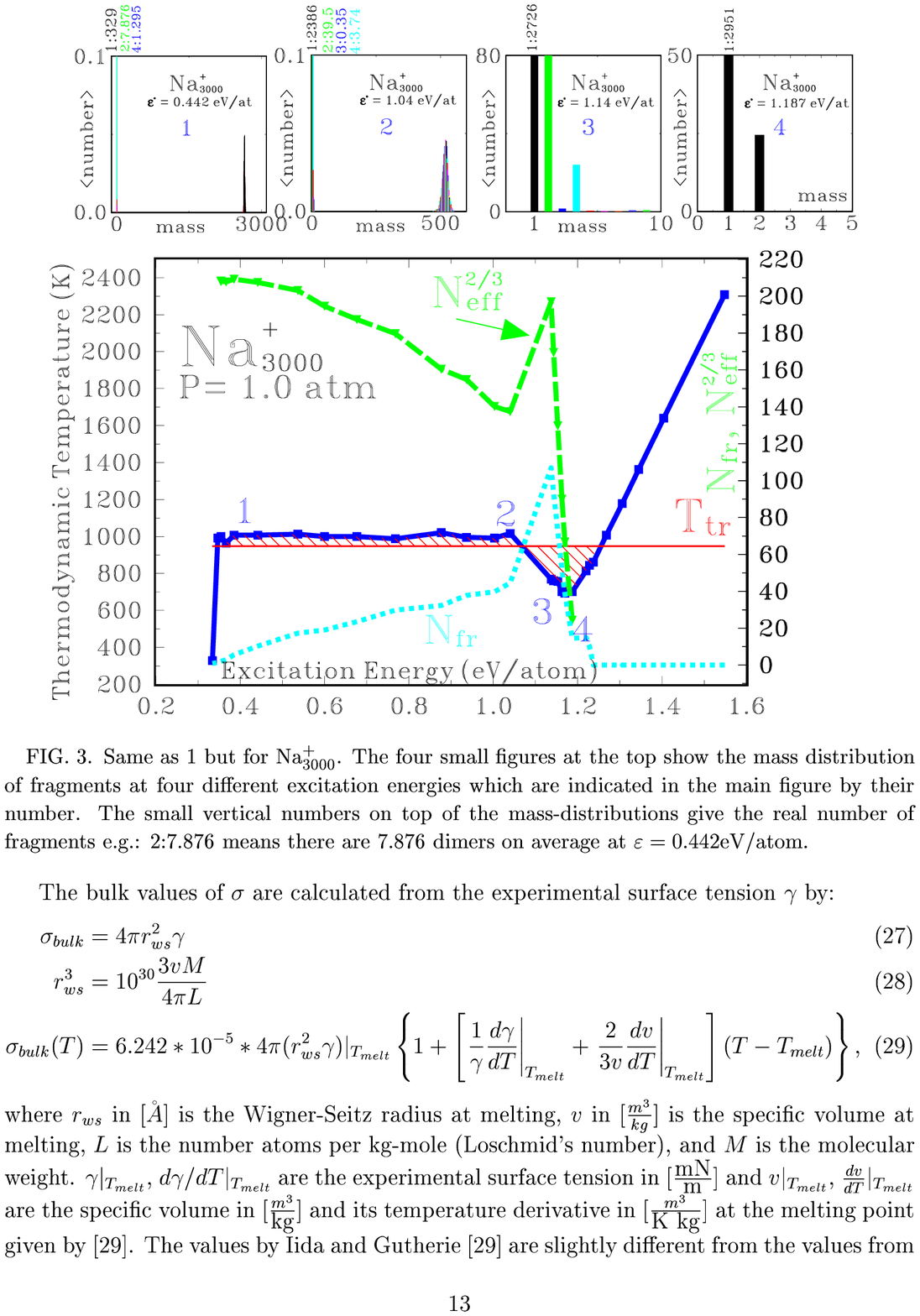}\caption{Cluster
fragmentation. Here I show the simulation of a typical
fragmentation transition of a system of $3000$ sodium atoms
interacting by realistic (many-body) forces. To compare with usual
macroscopic conditions, the calculations were done at each energy
using a volume $V(E)$ such that the microcanonical pressure
$P=\frac{\partial S}{\partial V}/\frac{\partial S}{\partial
E}=1$atm. The inserts above give the mass distribution at the
various}
\end{minipage}~~\begin{minipage}[h]{8cm}points. The label
"4:1.295" means 1.295 quadrimers on average. This gives a detailed
insight into what happens with rising excitation energy over the
transition region: At the beginning ($e^*\sim 0.442$ eV) the
liquid sodium drop evaporates 329 single atoms and 7.876 dimers
and 1.295 quadrimers on average. At energies $e>\sim 1$eV the drop
starts to fragment into several small droplets ("intermediate mass
fragments") e.g. at point 3: 2726 monomers,80 dimers,$\sim$10
trimers, $\sim$30 quadrimers and a few heavier ones up to 10-mers.
The evaporation residue disappears. This multifragmentation
finishes at point 4. It induces the strong backward swing of the
caloric curve $T(E)$. Above point 4 one has a gas of free monomers
and at the beginning a few dimers. This transition scenario has a
lot similarity with nuclear multifragmentation. It is also shown
how the total interphase surface area, proportional to
$N_{eff}^{2/3}=\sum_i N_i^{2/3}$ with $N_i\ge 2$ ($N_i$ the number
of atoms in the $i$th cluster) stays roughly constant up to point
3 even though the number of fragments ($N_{fr}=\sum_i$) is
monotonic rising with increasing excitation.
\end{minipage}
\end{figure}
\subsection{Stars} The appearence of a negative heat capacity of
equilibrized self gravitating systems is well known
\cite{thirring70}. It was always considered as an absurd pitfall
of thermodynamics. In our generalized theory this phenomenon turns
out as the standard occurrence of pseudo-Riemannian geometry of
$S(E,L,N)$ c.f.\cite{gross187}.
\begin{figure}[h]
\begin{minipage}[h]{8cm}

    \includegraphics*[bb =88 401 530 605, angle=-0, width=8 cm,
    clip=true]{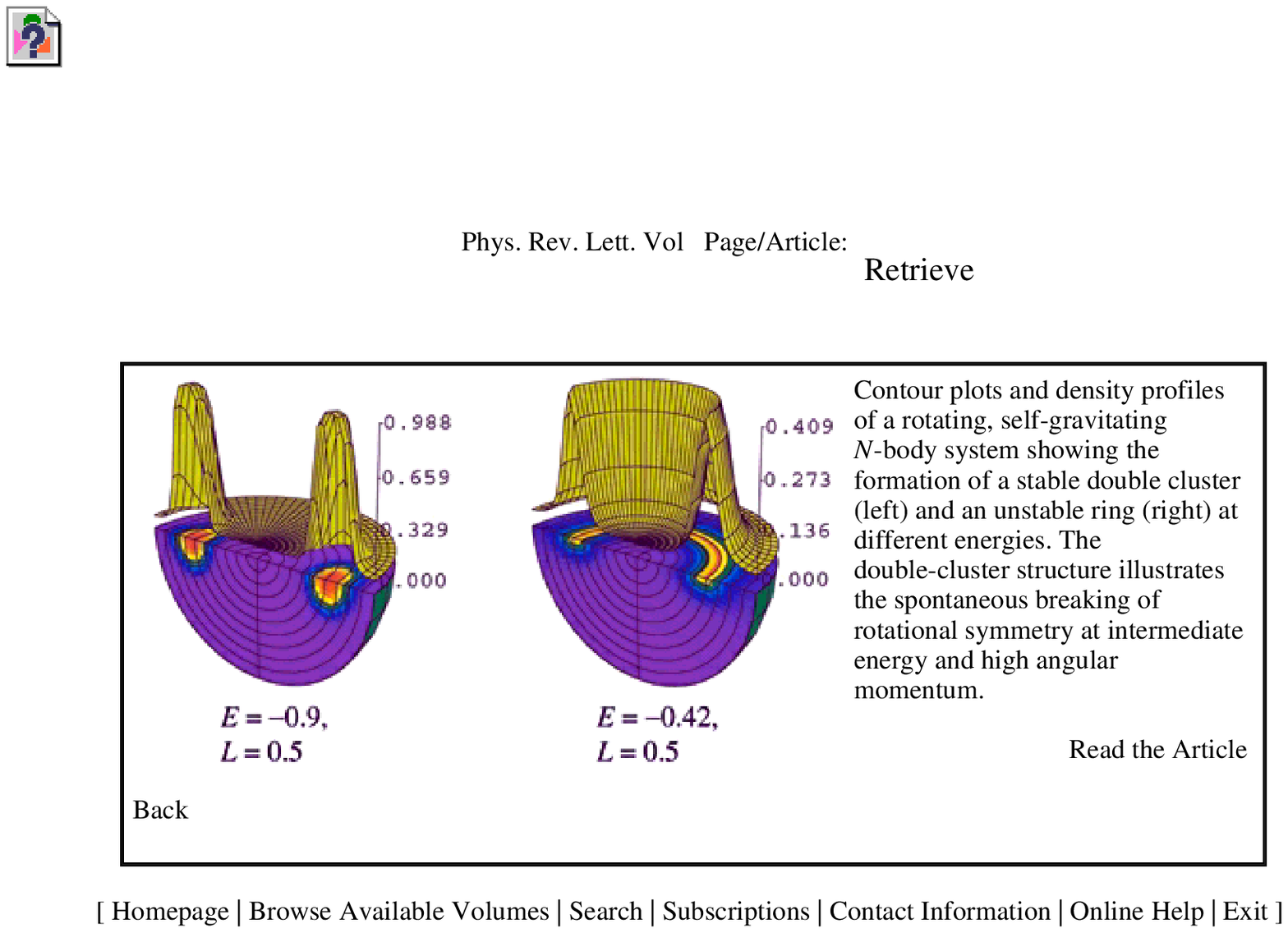}
    \caption{Phases and Phase-Separation
    in Rotating, Self-Gravitating Systems,
    Physical Review Letters--July 15, 2002, cover-page, by
    (Votyakov, Hidmi, De Martino, Gross}
\end{minipage}~~\begin{minipage}[h]{8cm}
%%    \end{figure}
%%\begin{figure}[h]
\includegraphics[bb =72 54 533 690,width=6cm,angle=-90,clip=true]{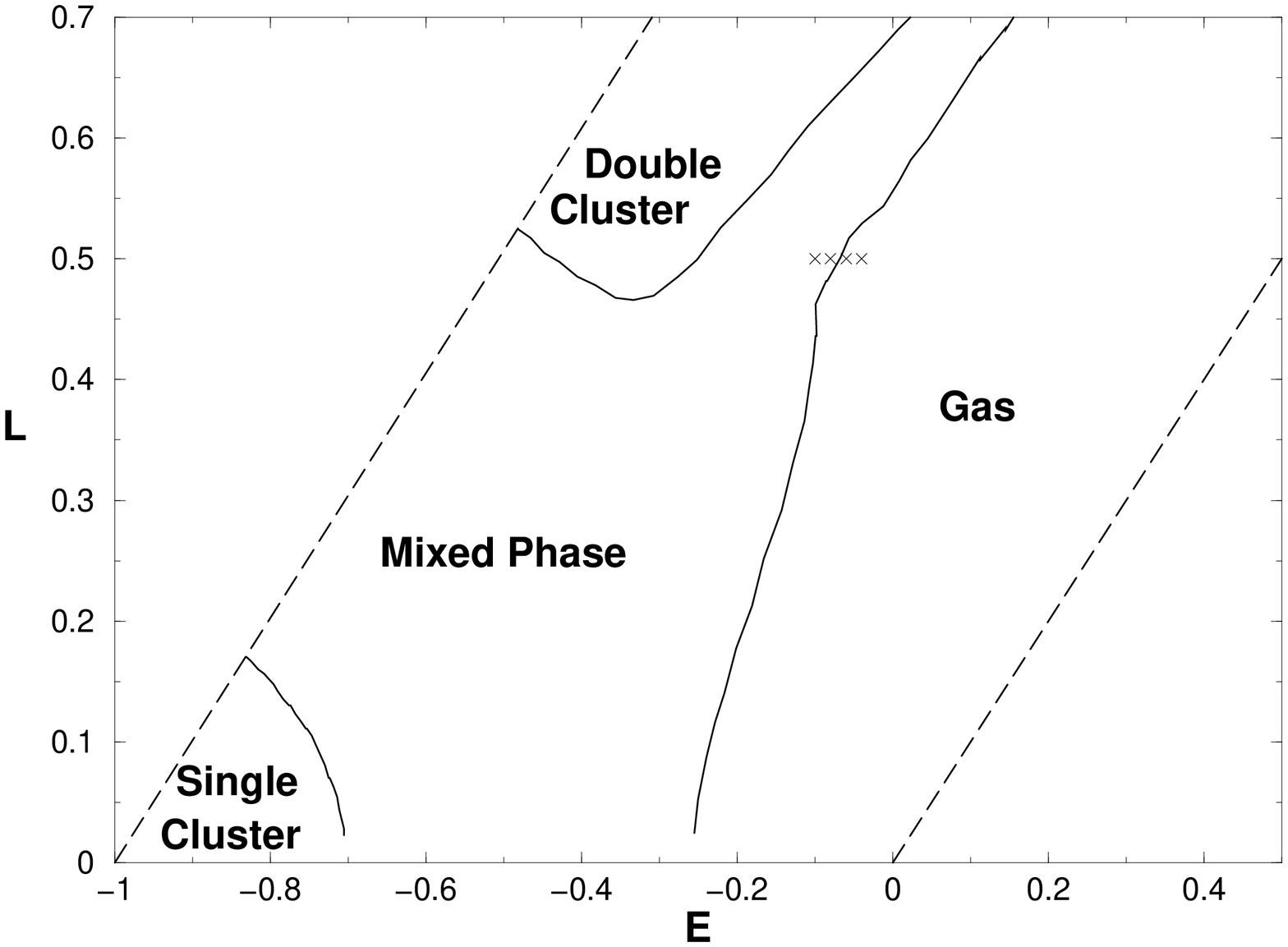}
\caption{ Microcanonical phase-diagram of a cloud of
self-gravitating and rotating system as function of the energy and
angular-momentum. Outside the dashed boundaries only some singular
points were calculated.}
\end{minipage}\end{figure}
 In the mixed phase the largest curvature
$\lambda_1$ of $S(E,L)$ is positive. Consequently the heat
capacity or the correspondent susceptibility is negative. The new
and important point of our finding is that within microcanonical
thermodynamics this is {\em a generic property of all phase
transitions of first order, whether there is a short- or a
long-range force that organizes the system}.

    Self-gravitation leads to a non-extensive potential energy $\propto
N^2$.  No thermodynamic limit exists for $E/N$ and no canonical treatment
makes sense. At negative total energies these systems have a negative heat
capacity.  This was for a long time considered as an absurd situation
within canonical statistical mechanics with its thermodynamic ``limit''.
However, within our geometric theory this is just a simple example of the
pseudo-Riemannian topology of the microcanonical entropy $S(E,N)$ provided
that high densities with their non-gravitational physics, like nuclear
hydrogen burning, are excluded. We treated the various phases of a
self-gravitating cloud of particles as function of the total energy and
angular momentum, c.f. the quoted paper. Clearly these are the most
important constraint in astrophysics.
\section{Conclusion}
Entropy has a simple and elementary mechanical definition by the
{\em area} $e^{S(E,N,\cdots)}$ of the microcanonical ensemble in
the $6N$ dim. phase space. Canonical ensembles are not equivalent
to the micro-ensemble in the most interesting situations:
\begin{enumerate}
\item At phase-separation (\lora heat engines !), one gets
 inhomgeneities, and a negative heat capacity or some other negative susceptibility,
\item  Heat can flow from cold to hot.
\item Phase transitions can be localized sharply and unambiguously in
small classical or quantum systems, there is no need for finite
size scaling to identify the transition.
\item Also really large self-gravitating systems can be addressed.
\end{enumerate}
Entropy rises during the approach to equilibrium, $\Delta S\ge 0$,
also for small mixing systems. i.e. the Second Law is valid even
for small systems \cite{gross183,gross192}. With this geometric
foundation thermo-statistics applies not only to extensive systems
but also to non-extensive ones which have no thermodynamic limit.

It is only by this extension of thermo-statistics that the real
and deeper sources of thermodynamics become uncovered. It
addresses a much larger world of phenomena. The old puzzle of the
anomalous behavior of self-gravitational systems is resolved. The
first time thermodynamics does justice to its original goal of
treating properly  the liquid-gas phase-separation, the motor of
steam engines.

%%%\newpage
%%%{\bf Some papers on Negative Heat in chronological order}
%%%\cite{thirring70,gross81,gross82,gross95,labastie90,gross99,gross124,gross124,lyndenbell95a,gross158,gross150,gross159,gross154,gross157,gross116,lyndenbell99,gross150,gross179,torcini98,antoni99,casetti99a,gulminelli99a,schmidt00,dAgostino00,gross124,schmidt01,gross183,gross174,gross176,tsallis01,leyvraz01,leyvraz01a,ispolatov01,gobet02,gross188,moretto01,gross179,moretto02,gross196,rapisarda02}
%%%\bibliographystyle{unsrt}%{alpha}%{plain} %{unsrt}
%%%\bibliography{gross,othbiba,othbibb,othbibcd,othbibe,othbibf,othbibg,othbibh,othbibij,othbibk,othbibl,othbibm,othbibn,othbibo,othbibp,othbibr,othbibs,othbibt,othbibuw,othbibxz}
%%%\end{document}

\end{document}